\def\year{2022}\relax
\documentclass[letterpaper]{article} 
\usepackage{aaai22}  
\usepackage{times}  
\usepackage{helvet}  
\usepackage{courier}  
\usepackage[hyphens]{url}  
\usepackage{graphicx} 
\usepackage{natbib}  
\usepackage{caption} 
\DeclareCaptionStyle{ruled}{labelfont=normalfont,labelsep=colon,strut=off} 
\usepackage{algorithm}
\usepackage{algorithmic}

\usepackage{newfloat}
\usepackage{listings}
\floatstyle{ruled}
\newfloat{listing}{tb}{lst}{}
\floatname{listing}{Listing}
\title{My Lockdown Escape: Sparking Self-Empathy in the Context of the Covid-19 Pandemic}

\author {
    Andrea Tocchetti,\textsuperscript{\rm 1}
     Silvia Maria Talenti, \textsuperscript{\rm 1}
     Marco Brambilla \textsuperscript{\rm 1}
}
\affiliations {
     \textsuperscript{\rm 1} Politecnico di Milano, Dipartimento di Elettronica, Informazione e Bioingegneria\\
     andrea.tocchetti@polimi.it, silviamaria.talenti@mail.polimi.it, marco.brambilla@polimi.it
}

\begin{document}

\maketitle

\begin{abstract}
During the Covid-19 pandemic, research communities focused on collecting and understanding people's behaviours and feelings to study and tackle the pandemic indirect effects. Despite its consequences are slowly starting to fade away, such an interest is still alive. In this article, we propose a hybrid, gamified, story-driven data collection approach to spark self-empathy, hence resurfacing people's past feelings. The game is designed to include a physical board, decks of cards, and a digital application. As the player plays through the game, they customize and escape from their lockdown room by completing statements and answering a series of questions that define their story. The decoration of the lockdown room and the storytelling-driven approach are targeted at sparking people's emotions and self-empathy towards their past selves. Ultimately, the proposed approach was proven effective in sparking and collecting feelings, while a few improvements are still necessary.
\end{abstract}

\section{Introduction}
The recent Covid-19 pandemic and its consequences affected our lives unprecedentedly, changing our daily habits whilst disrupting our emotional and psychological health \cite{Kontoangelos_2020}. Among these consequences, the lockdown enforced by the local governments caused some of the most devastating ones, including depression, anxiety, and stress \cite{Fiorillo_2020, Armbruster_2020, Elhadi_2021}. Although the pandemic's consequences are slowly fading away, the research community's interest in understanding people's emotions over that period is still alive \cite{Tan_2023}. Researchers with different backgrounds have been shaping data collection processes and developing methodologies to involve people in sharing their feelings by designing various approaches combining gamification techniques with the most commonly employed survey approach \cite{Rabbi_2017, Triantoro_2020}.
While these methods may still be effective regardless of time, people are slowly starting to forget how they felt. For this reason, collecting people's feelings requires designing approaches capable of sparking these emotions again. 

This article proposes a gamified approach to collect people's feelings during the Covid-19 pandemic named "My Lockdown Escape". The proposed methodology strives to make people empathise with their past selves -- a concept we call \textit{self-empathy} -- by combining different gamified design techniques. The proposed design follows a hybrid approach with both digital and physical elements. A digital application implementing a storytelling-driven activity supports an escape room-like experience involving decks of cards and a board. In this article, we strive to demonstrate the effectiveness of the proposed methodology in collecting people's feelings by sparking their empathy towards their past selves. Furthermore, we report on the collected data and discuss interesting insights into people's perspectives and feelings towards their experience during the pandemic.

The remainder of this article is organized as follows. Chapter 2 describes empathy and gamification in the context of interest. Chapter 3 describes the design and implementation of the proposed method, focusing on the first. Chapter 4 reports on the structure of the experiments and the profiles of the participants. Chapter 5 discusses the results, user experience, and the analysis of the collected data. Chapter 6 summarises the work and provides some insights into future works and design improvements.

\section{Related Works \& Background}
\subsection{Empathy \& Self-Empathy}
Empathy can be described as \textit{the capability of a human to put themselves in someone else's shoes} \cite{Hardee_2003}. An extensive definition characterises empathy as \textit{an emotional response, dependent upon the interaction between trait capacities and state influences, whose resulting emotion is similar to one’s perception (directly experienced or imagined) and understanding (cognitive empathy) of the stimulus emotion, with the recognition that the source of the emotion is not one’s own} \cite{Cuff_2016}.
Regardless of the difference in complexity, these definitions imply a social relation between two people: the one who feels and expresses an emotion and the one who experiences the consequent emotional response.
We stray from such a standard model of empathy, focusing on sparking and assessing people's empathy compared to their past selves rather than someone else, resulting in a so-called one-state model \cite{Goldie_2011}. Such a change of perspective should drive the person towards a better understanding of the experienced emotions since they were the ones feeling them in the first place.

The research field on empathy found fertile ground in computer science \cite{Wright_2008}, resulting in the development and assessment of various approaches leveraging gameful design elements to drive empathy \cite{LopezFaican_2021, Kors_2016, PeaceMaker}. 
Such a combination of the physical and digital worlds makes it necessary to highlight a fundamental difference between the empathy experienced by humans through their peers and the one conveyed through digital technologies. The first is a human reaction sparked by our perception and understanding of the feelings of another human being through our senses. On the other hand, the second must leverage digital technologies' features to spark it. In particular, they rely on images and sounds, like photos \cite{Mauri_2022}, videos \cite{Pantti_2013, Tettegah_2016}, music \cite{Clarke_2015}, etc., to convey emotions, feelings, and perceptions since digital technologies lack the ability to convey them through the senses. Hence, it is necessary to design approaches capable of dealing with such a gap, driving people to empathise even when digital environments are employed.

\subsection{Gamification}
Gamification can be defined as the application of game design elements in non-game contexts to invoke gameful experiences and behavioural outcomes from users to support the value of the content they provide or create \cite{Huotari_2017, Hamari_2014}. Such an approach has been applied to several fields, \textit{e.g.}, medical and healthcare \cite{Allam_2015}, policy-making \cite{Mauri_2022, Tocchetti_2020, Tocchetti_2021}, educational \cite{Lam_2022, Nieto_2021} and many more, demonstrating the all-around applicability of such a paradigm.
The effects of such approaches have been studied for a very long time by the research community. For example, classic gamification elements (\textit{e.g.}, avatars, animations, challenges, etc.) were proven to be effective in improving user attention and enjoyment \cite{Triantoro_2020}, improving participation rate and reliability of the answers in data collection tasks \cite{Steinmaurer_2021}, as well as improving user engagement and driving behaviours \cite{Ahmed_2021}.

The Covid-19 outbreak's consequences heavily impacted people's emotional and psychological health, sparking research to analyse the pandemic's scale and impact \cite{Islam_2020, Cellini_2020, Larcher_2021, Hung_2020, Philpot_2021}. In that regard, researchers developed gamification strategies to achieve better coverage and user involvement whilst delivering an interesting and enjoyable experience \cite{Mauri_2022, Tocchetti_2020, Tocchetti_2021}.
While understanding people's emotional and psychological conditions has been fundamental to comprehending the impacts of the pandemic, some researchers focused on tackling its consequences, demonstrating the effectiveness of gamified approaches in motivating and enhancing students' learning \cite{Videnovik_2022, Gerber_2022}, approaching elderly people with healthcare initiatives \cite{White_2022}, improving the population's awareness about disinformation \cite{Coronavirus_Quiz, Go_Viral}, etc. Among these gamified activities, some applied specific design elements to achieve their objectives. In particular, escape room-like experiences were proven effective in springing cooperation \cite{Gerber_2022} and motivation \cite{Gerber_2022, Gomez_2019, Sanchez_2020, Videnovik_2022} among participants, especially when applied in remote digital environments. On the other hand, digital storytelling (\textit{e.g.}, narratives, interactive stories, etc.) was shown to improve the application's appealing \cite{Giakalaras_2016}, user engagement \cite{Molnar_2018}, and rising emotions and sparking imagination \cite{Cesario_2019} as they engage the user on a personal level in a novel or familiar experiences. Despite its demonstrated effectiveness, the research community still acknowledged the need to apply gamification carefully (\textit{e.g.}, avoid biasing the user with the narrative \cite{Harteveld_2018}, or avoid using reward-based mechanisms in surveys \cite{Steinmaurer_2021}) to prevent undesired outcomes and/or behaviours.

\begin{figure*}
    \centering
    \includegraphics[width=\textwidth]{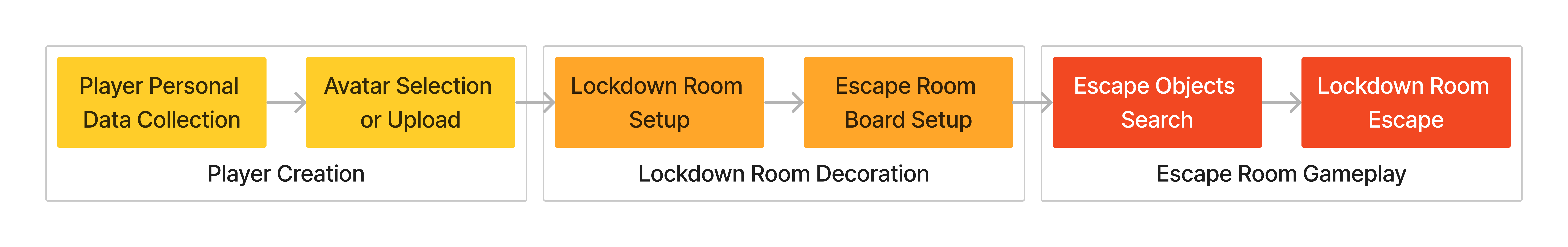}
    \caption{A high-level representation of the three steps of the game, namely Player Creation, Lockdown Room Decoration, and Escape Room Gameplay, and their sub-steps.}
    \label{fig:process}
\end{figure*}

\section{Method}
Acknowledged the advantages and weaknesses of the previously discussed gamified techniques, we strive to combine them to design an activity called "My Lockdown Escape" to spark emotions and drive people to self-empathise with their past selves, finally moving them to describe and share the feelings they experienced during the Covid-19 pandemic. The approach is designed to be hybrid, tackling the drawbacks of a fully digital design in sparking empathy whilst acknowledging its advantages (\textit{e.g.}, ease of use, process automation, etc.). The activity is designed as a story-driven escape room experience through which the player describes the room they spent their lockdown in and interacts with some items of fundamental usage during the pandemic to prepare themselves to leave their room. Decks of cards are used to represent the different categories of items the player decorates their house and interacts with. A board with card slots is used to model the environment where their lockdown experience occurred. An interactive digital application guides the player throughout the activity by means of a storytelling-based approach that provides the player with a series of situations they have (most likely) experienced during the pandemic. Such an application is optimised to run on mobile devices and use their features (\textit{e.g.}, camera). The combination of these elements and their design lead the player to describe the mental and emotional conditions they experienced during the lockdown. In particular, we argue that having the player remember the room they lived in would spark memories and feelings from their lockdown experience. Furthermore, the similarity between the context and the considered design elements (\textit{i.e.}, the escape room design and the lockdown, and the storytelling-driven design and the people's experience) makes the final activity very close to the real experience, hence contributing to sparking emotions and stimulating self-empathy.

The proposed activity can be divided into three main steps (represented in Figure \ref{fig:process}): 
\begin{itemize}
    \item \textbf{Player Creation}, \textit{i.e.}, collecting the player's personal data and customising their avatar,
    \item \textbf{Lockdown Room Decoration}, \textit{i.e.}, decorating the lockdown room by placing the cards on the board and setting up the next step of the activity,
    \item \textbf{Escape Room Gameplay}, \textit{i.e.}, escaping from the lockdown room by collecting the cards placed on the board. \\
\end{itemize}

\textbf{Player Creation} -- The player is asked to provide their personal data through the digital application. They provide a nickname to guarantee anonymous data collection, age, country, gender, ethnicity, and education level. They are also requested to create a physical avatar or pick a digital one. In the first case, they customize their avatar card, \textit{i.e.}, a card with a simple outline of a stylized person, by drawing on it using coloured markers. Then, such a card is uploaded into the system by taking a picture and placed in the corresponding slot on the board. Alternatively, the player can pick one of the pre-made avatars available on the digital application. If a digital avatar is chosen, the corresponding physical avatar card is positioned on the board. \newline

\begin{figure*}
    \centering
    \includegraphics[width=0.8\textwidth]{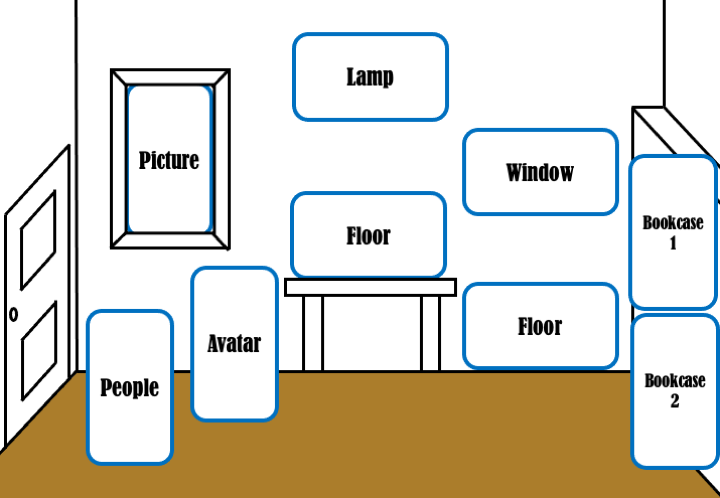}
    \caption{A representation of the part of the board to support the Lockdown Room Decoration step. Each slot has an associated deck name that represents the deck from which the card to be placed belongs. An avatar slot where the player can place their avatar card is also featured. In the considered setting, all decks have one slot each, besides the Floor deck that has two.}
    \label{fig:board_1}
\end{figure*}

\begin{figure}[H]
    \centering
    \includegraphics[width=0.5\textwidth]{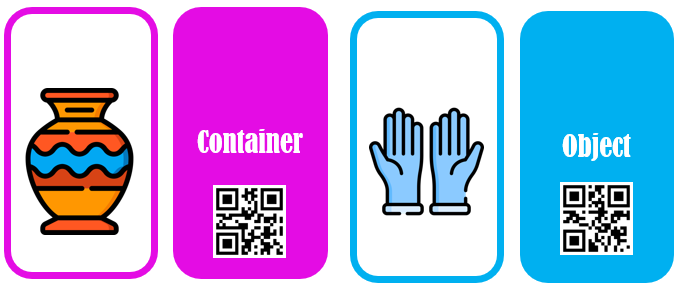}
    \caption{Examples of cards from the decks involved in the Escape Room Gameplay step (\textit{i.e.}, \textit{Container} on the left and \textit{Object} on the right).}
    \label{fig:cards}
\end{figure}

\textbf{Lockdown Room Decoration} -- The player decorates their lockdown room by completing statements in the story narrated through the digital application. Each statement is assigned to a part of the story and a deck whose cards represent a possible phrase or word to complete the sentence. Decks are identified by colour and a unique name. Every card has a symbol printed on its front and a unique QR code and the name of the deck they belong to on its back. Examples of cards are depicted in Figure \ref{fig:cards}, on the left. At this stage, the part of the board to be used (represented in Figure \ref{fig:board_1}) represents the lockdown room, \textit{i.e.}, an abstraction of a real room where the player experienced the pandemic. It has at least one dedicated card slot for each of the seven decks involved, namely \textit{People}, \textit{Picture}, \textit{Floor}, \textit{Lamp}, \textit{Window}, \textit{Bookcase 1}, and \textit{Bookcase 2} which represent the elements the player uses to describe their room.
\newline
For each statement, the player inspects and picks a card of choice from the associated deck, scans the QR code on its back using their mobile phone through the application, and places it face-up in the corresponding board slot. Scanning the QR code stores the choice in the system. Whenever a statement is completed, the corresponding part of the story is updated and displayed alongside the next one. Such a process is repeated until a card is placed in all the slots associated with the seven involved decks. An example of a statement and the corresponding list of completions are provided below.

\begin{quote}
\textit{\textbf{Statement:} Our story begins in early 2020, and not long ago, the COVID-19 pandemic broke out in your country and the whole world. You are at home watching the news. The titles are scary and doubtful. Take a look around you, and you will see that you are surrounded by ...}
\end{quote}

\begin{figure}
    \centering
    \includegraphics[width=0.5\textwidth]{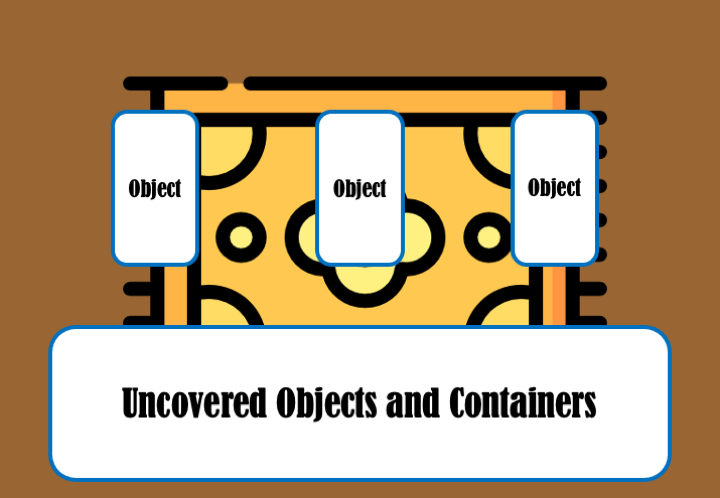}
    \caption{A representation of the part of the board to support the Escape Room Gameplay step. The three slots on top are dedicated to the three piles of cards that will be prepared from the \textit{Object} and \textit{Container} decks, while the cards uncovered by the player will be placed in the slot at the bottom.}
    \label{fig:board_2}
\end{figure}

\begin{quote}
\textit{\textbf{Possible Completions (\textit{i.e.}, Cards):} Family, Parents, Friends, Strangers, No one, Roommates, and Animals} 
\end{quote}
\ \newline
Then, the player setups the board to be used in the second part of the activity (represented in Figure \ref{fig:board_2} on the right). They shuffle the \textit{Object} deck and create three face-down piles to be placed on three dedicated slots on the board by evenly distributing the cards. Then, one randomly selected \textit{Container} card is placed atop each pile. While the \textit{Object} cards represent the different items the person could have interacted with during the pandemic, the \textit{Container} cards represent the furniture in which they are stashed. \newline

\textbf{Escape Room Gameplay} -- The player must now find three core \textit{Object} cards, \textit{i.e.}, the mask, hand sanitiser, and green pass cards, to escape their lockdown room. They must scan the QR code on the back of the card on top of a pile of choices and answer the corresponding question in the digital application. An example of a question and the corresponding list of answers are provided below.

\begin{quote}
\textit{\textbf{Question (\textit{i.e.}, Cards):} Think back at your lockdown experience. If it was a movie, what title would it have?}
\end{quote}

\begin{quote}
\textit{\textbf{Possible Answers:} The Never-ending Story, The Social Network, Home Alone, Life is Beautiful, Back to the Future, Eat Pray Love, A Good Year, Cast Away}
\end{quote}
\ \newline
When a question is successfully answered, the player flips its card and uncovers the item it hides. Whenever an item is discovered, its card is placed in the dedicated area of the board (represented in Figure \ref{fig:board_2}, on the right) alongside all the other items the player has already found and its corresponding digital icon is displayed in the application. Such a process is the same regardless of whether the item belongs to the \textit{Object} or \textit{Container} deck. At this stage, the board represents the spot (\textit{e.g.}, a carpet, a table, etc.) where the player places the items they uncover. This process is repeated until all the \textit{Object} cards necessary to escape have been found. Then, the player can escape the room or keep playing to find all the objects. \newline
When they successfully escape their lockdown room, the player is shown their story which can then be shared with their peers. In particular, it includes a picture of the player's avatar or their virtual representation, the textual description of the room they decorated with the completed statements, and the items they uncovered with the corresponding questions' answers. \newline

"My Lockdown Escape" is designed following a hybrid setting, combining physical and digital assets. The physical assets (\textit{i.e.}, the cards and the board) were designed using digital tools. Then, they were printed on cardboard, cut, and coated with plastic. The digital asset (\textit{i.e.}, the web application) was developed abiding by the structure of a three-layer architecture. The front end was implemented using HTML, CSS, Javascript, and Thymeleaf. Furthermore, the Bootstrap toolkit was widely employed. The middle layer was developed using Java, Spring Boot, and the Model-View-Controller framework. The back end is managed through a relational database implemented using MySQL technology. Such an application was deployed on a web server to make it accessible to multiple players simultaneously.

\section{Experiments}
We evaluated the effectiveness of the proposed approach in a series of experiments with different objectives.
The first experiment involved 21 students and researchers (9 women and 12 men) from an Italian university, mainly aged between 21 and 27 years old (26,7 years old on average), in a series of individual experiments in Milan. The second one involved 28 people (17 women and 11 men) from a variety of European organizations, mainly aged between 22 and 66 years old (28,4 years old on average), in an open experiment in Bruxelles. Whilst the first experiments was mainly aimed at collecting feedback about the approach and the user experience, the second one contributed to test the methodology in an open environment and collect feedback about possible improvements. The participants to both experiments were given an initial description of the application. Then, they performed the activity without receiving any suggestions. Each participant was required to bring their own mobile phone to play. Such a setting allowed the testing of the application on different mobile operative systems and web browsers.

As previously described, the approach was mainly designed to collect the participants' feelings by sparking self-empathy in the context of the Covid-19 pandemic. The questions the participants answered and the statements they completed were aimed at collecting such data. Regarding the achievement of this objective, we recognize the nondeterministic nature of the data collection performed in the Escape Room Gameplay step. Indeed, the player may escape the room before answering to all the questions after they found the three \textit{Object} cards. We argue that such an event does not impact the assessment of our method as it only influences the amount of data collected in the dedicated part of the game. Moreover, it would be quite easy to reshape the rules of the game to have the player answering all the questions before escaping the room, \textit{e.g.}, by allowing the player to leave only after the three piles are empty.
To assess the effectiveness of our methodology, each of the first experiment's participant was asked to answer a questionnaire including all the questions from the System Usability Scale (SUS) \cite{Brooke_1996} (10 questions) to measure the system's usability \cite{Bangor_2009}, a set of questions to evaluate the overall approach (inspired by \cite{Junior_2021}) (5 questions), and a set of questions to evaluate the tool's effectiveness in sparking self-empathy (inspired by GEQ \cite{IJsselsteijn_2013} and GUESS \cite{Phan_2016}) (5 questions). The latter were custom-made since there are very few or no questionnaires addressing the assessment of self-empathy and hybrid approaches in the literature. A list of such questions is available in Appendix A.
The questions' order in the questionnaire was randomized to prevent potential bias. The answers were modelled following a Likert Scale approach ranging from 1 ("Strongly Disagree") to 5 ("Strongly Agree"). 

\section{Results}
\subsection{Approach Assessment}
Ultimately, the first experiment yielded positive results and provided useful feedback to improve the approach. In particular, the application achieved a final SUS score of 75 which represents good usability compared to the average SUS score of 68 \cite{MeasuringUsability}.
We achieved a score of 78\% and 72\% for the hybrid approach and empathy assessment, respectively, by averaging the numerical values on the corresponding Likert Scale of the answers. These scores provide preliminary evidence that the hybrid design is appreciated and the approach can spark empathy in most participants.
Despite most participants deemed the experience to be enjoyable and engaging, from the feedback we received, the behaviours we observed, and the computed scores, we acknowledge there's still room for improvement. First, the game's instructions may benefit from small clarifications and extra details. In particular, in the Lockdown Room Decoration step, some participants were mislead to take their cards randomly instead of picking them. Such a misunderstanding also caused them to position their cards face-down on the board instead of face-up.
Furthermore, when comparing the game steps, participants preferred the Lockdown Room Decoration step, stating that the Escape Room Gameplay may benefit from a small re-design due to the randomness in finding the cards to meet the escape condition. Additionally, we noticed that one of the most common behaviours was that most participants tended to leave the room as soon as they met the escape conditions. As previously discussed, even a slight change to the rules would allow to prevent such behaviour, finally improving the data collection. Regarding the latter objective, a few participants also stated that while the Lockdown Room Decoration step perfectly masked the data collection, they perceived it clearly in the Escape Room Gameplay step. Such feedback calls for improvements to better bind the approach with the data collection activity underneath. We argue that a better alignment between the cards and the associated questions would address such a drawback.

\subsection{Data Collection}
In this section, some of the insights we derived from our analysis are reported, mainly focusing on some statements from the Lockdown Room Decoration step and some questions from the Escape Room Gameplay step.

\begin{figure*}
    \centering
    \includegraphics[width=0.49\textwidth]{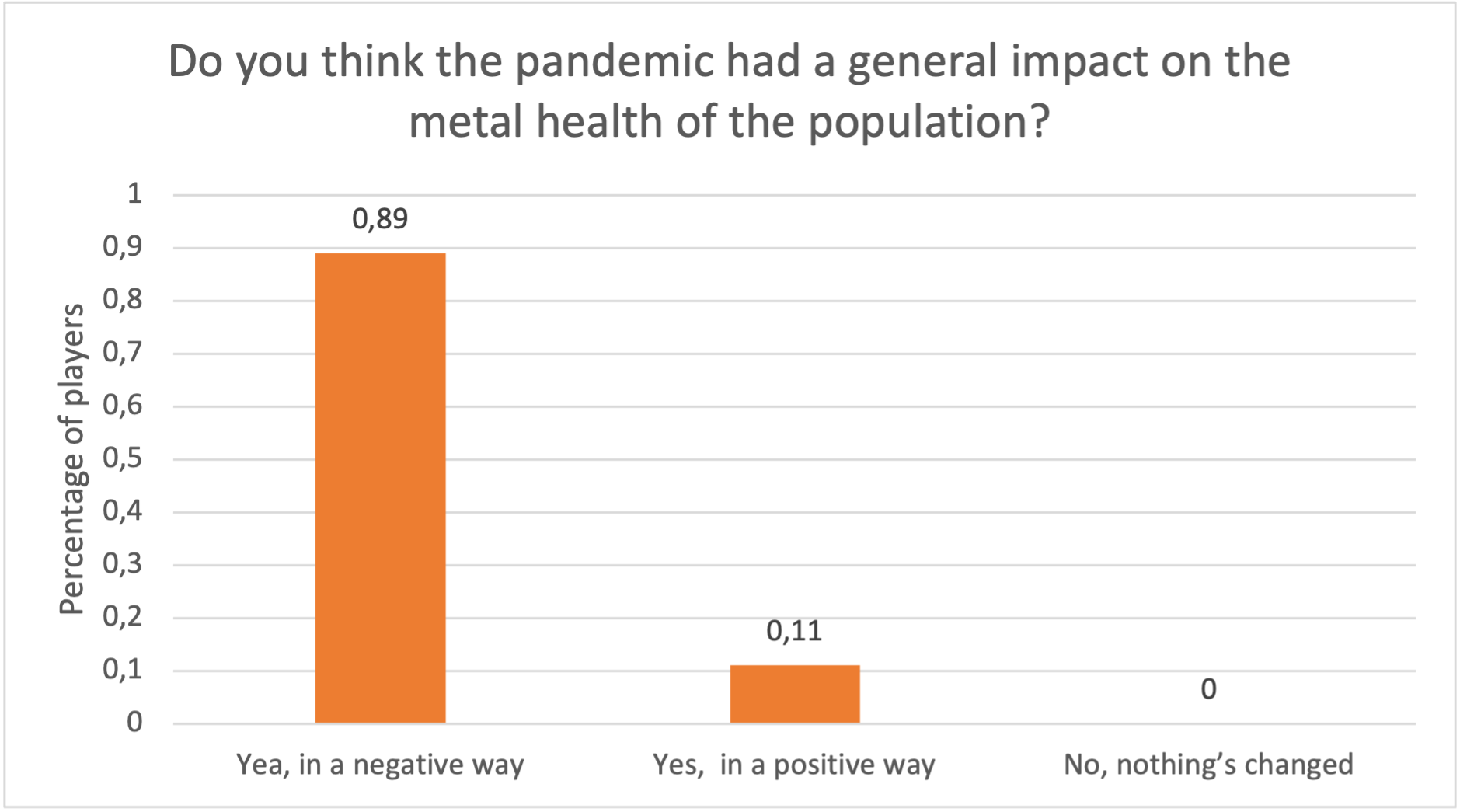}
    \hspace{0.01\textwidth}
    \includegraphics[width=0.49\textwidth]{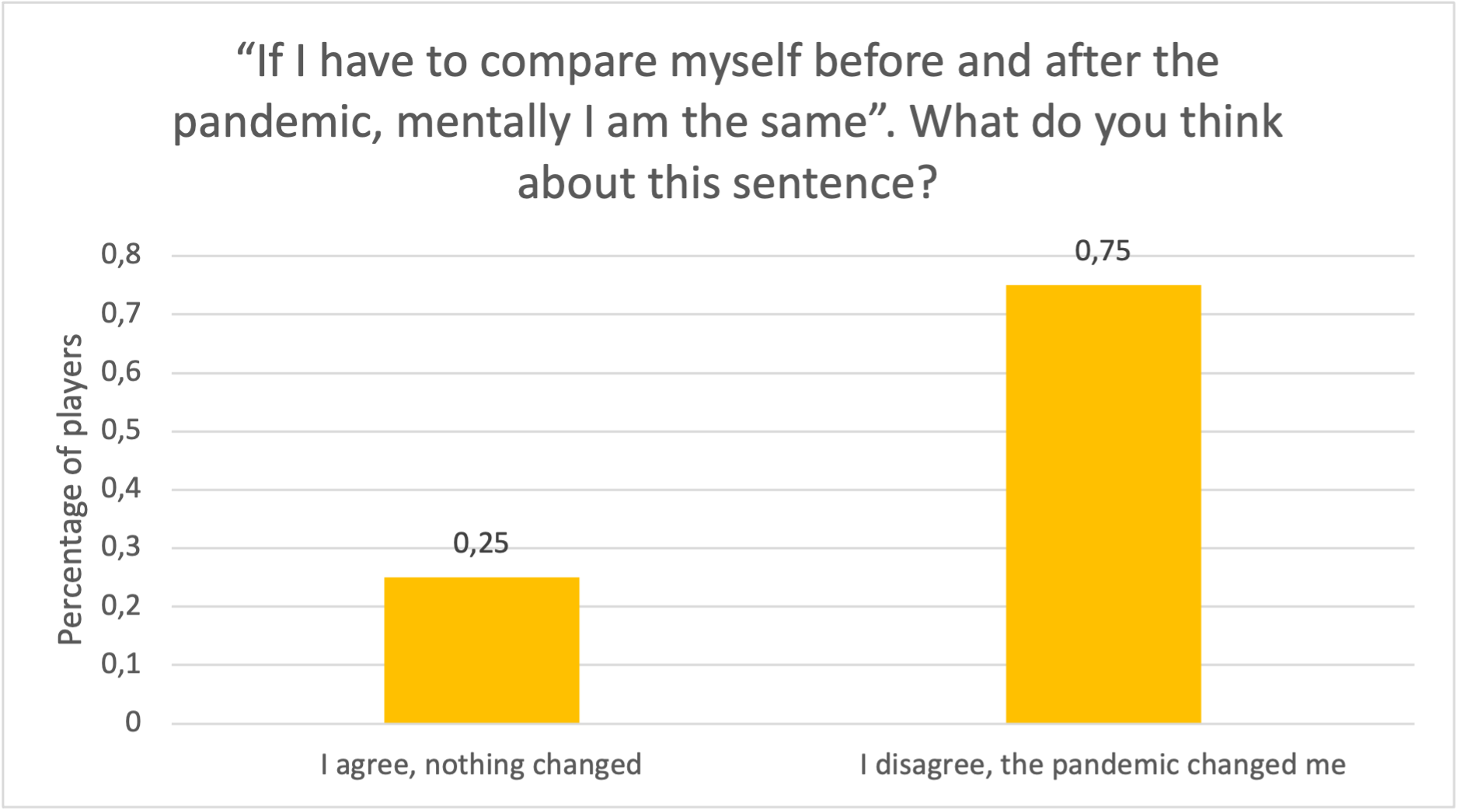}
    \caption{On the left,  distribution of answers to the question "Do you think the pandemic had a general impact on the mental health of the population?". On the right, percentages of participants that agreed or disagreed on whether the pandemic changed their mental health.}
    \label{fig:impact_vs_mentally}
\end{figure*}

At first, the collected data confirmed an obvious trend, \textit{i.e.}, most participants (89\%) think the pandemic negatively impacted the mental health of the population (as represented in Figure \ref{fig:impact_vs_mentally} on the left), while surprisingly revealing that a fair percentage (25\%) of the participants was not influenced at all (as represented in Figure \ref{fig:impact_vs_mentally} on the right). Similar trends can be identified in other statements. For example, regarding the statement "The enforcement of the lockdown made me feel...", most of the participants completed it using the words "Frustrated" (39\%) and "Anxious" (26\%), hence highlighting the negative impact of the lockdown on their mental and emotional health. On the other hand, a few participants used positive (less than 10\%) or neutral (less than 10\%) feelings, showing that not everybody was negatively impacted by the lockdown.
In the question "Think back at your lockdown experience. If it was a movie, what title would it have?", the titles that got chosen the most (more than 90\%) are the ones sparking negative emotions (\textit{e.g.}, "The Never Ending Story", "A Quiet Place", "Home Alone", etc.), once again confirming the general trend of negativity associated with the lockdown. A similar trend was also identified in the questions "You encounter a friend while walking on the street. He does not keep a correct social distance. How does it make you feel?" and "A friend of yours calls you to hang out at their place with other friends. Would you go?". In particular, we identified a general trend of negativity towards interacting with other people, even when friends are involved. Indeed, most participants (69\%) won't leave or would be afraid to leave their house to engage in social interactions while most participants would feel "Anxious" (30\%) or "Vulnerable" (30\%) when approached by someone who doesn't maintain the correct social distance.

\begin{figure}
    \centering
    \includegraphics[width=0.5\textwidth]{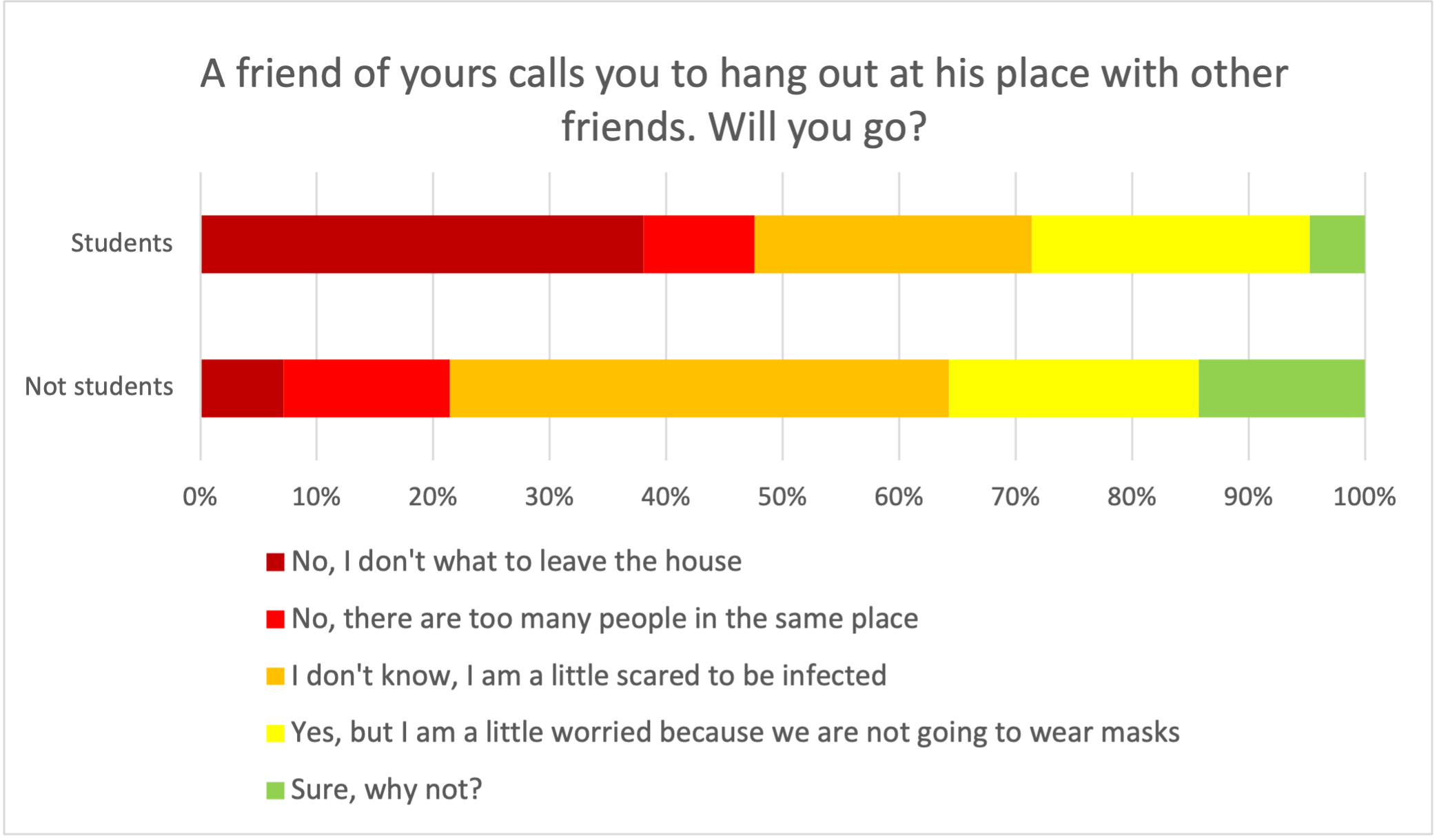}
    \caption{Answers to the question "A friend of yours calls you to hang out at their place with other friends. Would you go?" divided by the participants' student status (\textit{i.e.}, student vs. non-student).}
    \label{fig:interactions}
\end{figure}

\begin{figure}
    \centering
    \includegraphics[width=0.5\textwidth]{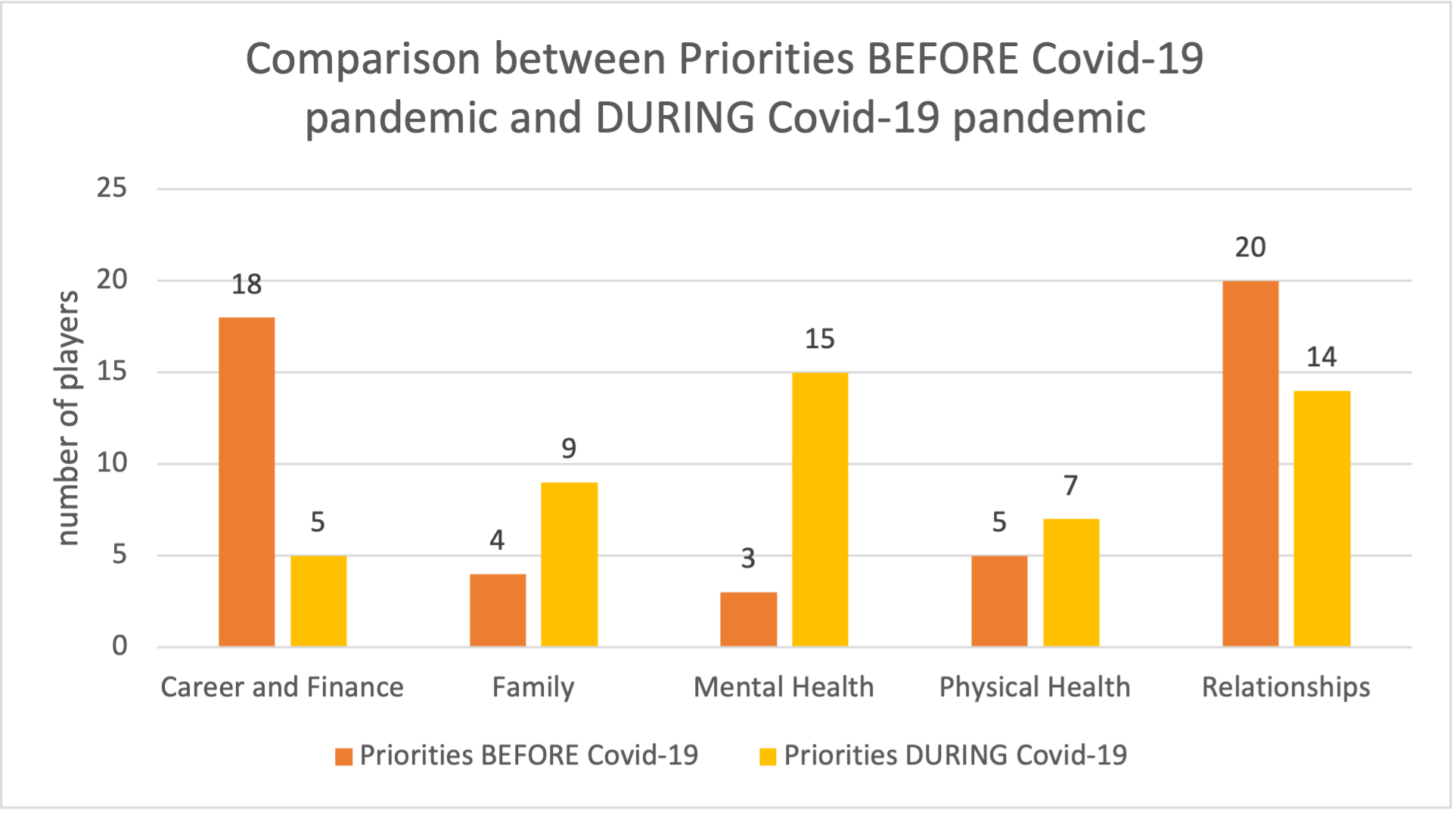}
    \caption{Distribution of the personal priorities of respondents before and during the pandemic.}
    \label{fig:priorities}
\end{figure}

Another interesting insight we observed is that the interests and priorities of the participants changed after the spread of the pandemic. In particular, most participants stated they were mainly interested in "Relationships" (40\%) and "Career and Finance" (36\%) before the pandemic. On the other hand, during the Covid-19 pandemic, their interests shifted towards "Mental Health" (30\%), "Relationships" (28\%) and "Family" (18\%).

\begin{figure}
    \centering
    \includegraphics[width=0.5\textwidth]{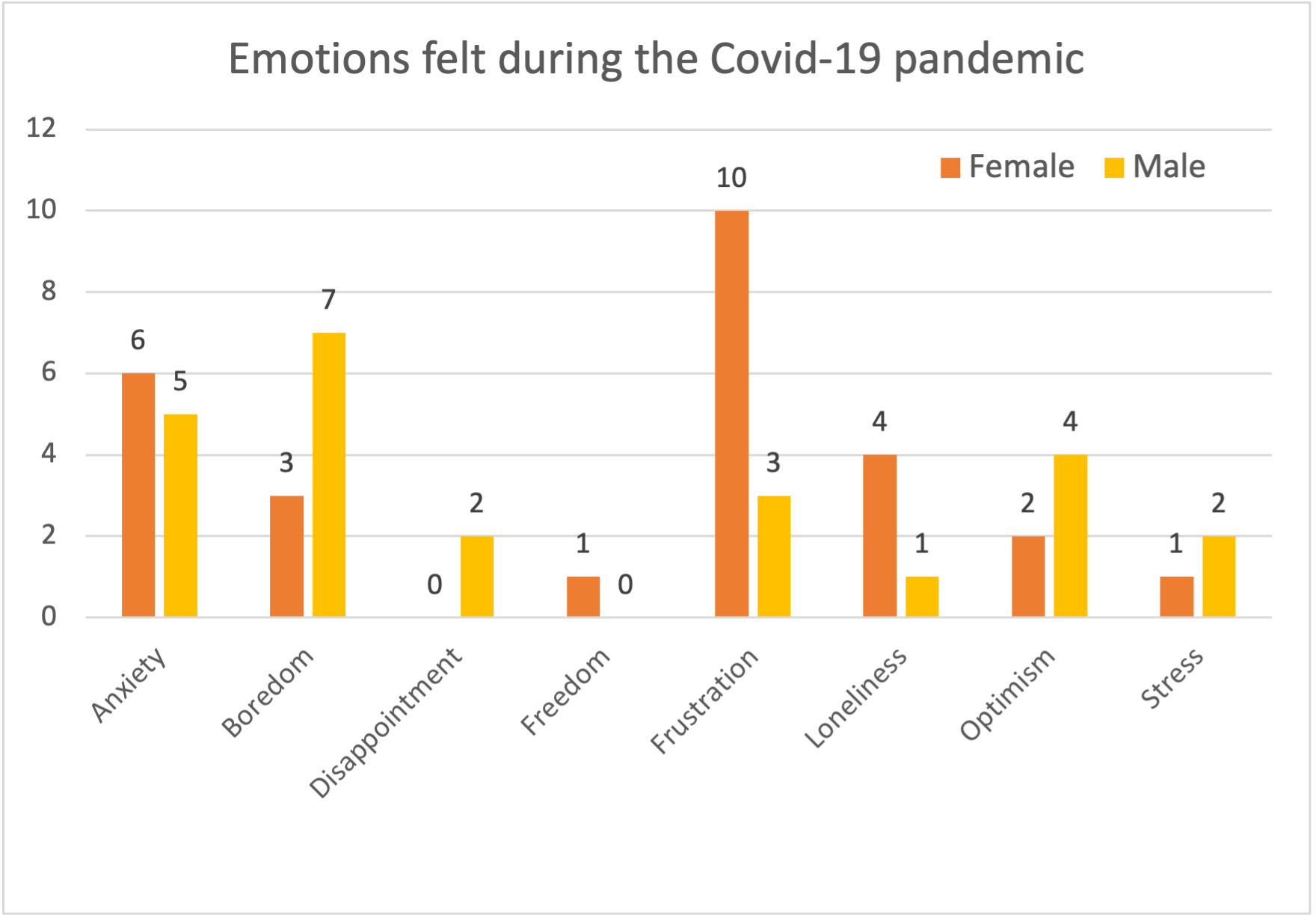}
    \caption{Distribution of the answers to "The main emotion felt during the Covid-19 pandemic" by gender (only male and female are represented since no participant picked other options).}
    \label{fig:gender}
\end{figure}

We also performed a few analyses based on their belonging to a specific group (\textit{i.e.}, students and non-students) for the last statement or gender (\textit{i.e.}, male or female) for the fourth question. In the first case (Figure \ref{fig:interactions}), we analysed how students and non-students would behave when social interactions are involved, identifying a stronger aversion in students to engaging in social activities. On the other hand, we identified trends in the feelings experienced during the pandemics based on gender (Figure \ref{fig:gender}). Indeed, while women mostly felt "Frustration", men mostly felt "Boredom", and both groups equally experienced "Anxiety". Such a result highlights that gender could be of fundamental interest when discerning the impacts and behaviours driven by the lockdown.

\section{Conclusions \& Future Works}
This article described a hybrid, gamified, story-driven data collection approach to spark self-empathy in participants. As they play and build their own story, they are driven to self-empathise with their past selves and provide data to be analysed to understand their past behaviours and attitudes. Preliminary experiments validated the approach and highlighted the need for a few improvements. 
In future works, we plan to improve the proposed gamified approach by addressing the feedback we received and providing new decks of cards and room abstractions, allowing even more freedom and customizability of the gameplay and the data to be collected.
Furthermore, we noticed that a small improvement could be factored into the game by shuffling the cards placed on the first part of the board with the \textit{Object} cards used to build the piles for the second part of the game. Such a change would improve the data collection while making the two parts of the game even more entwined.

\appendix
\section{Appendix A: Questionnaire}
\label{appendix:questionnaire}
The following questions were employed to assess the validity of the approach in our experiments.

\begin{itemize}
    \item I found this hybrid method more engaging than digital-only methods.
    \item I feel that this hybrid method is better than full-digital or full-physical.
    \item The escape-room style helped me remember my lockdown experience.
    \item The storytelling style helped me remember my lockdown experience.
    \item I found the game boring.
\end{itemize}

The following questions were employed to assess the capability of the approach to sparking self-empathy in our experiments.

\begin{itemize}
    \item The game helped me remember my lockdown experience.
    \item I would describe myself as a pretty soft-hearted person.
    \item When I think about sad past events of my life, I feel the same sadness.
    \item I am often quite touched by things that I see happen.
    \item The game helped empathize with my past self.
\end{itemize}

\section{Acknowledgements and Credits}
Container (Figure \ref{fig:cards} on the left) icon by Smashicons from www.flaticon.com
\newline
Object (Figure \ref{fig:cards} on the left) and Carpet (Figure \ref{fig:board_2}) icon by Freepik from www.flaticon.com

\bibliography{aaai22}

\begin{thebibliography}{49}
\providecommand{\natexlab}[1]{#1}

\bibitem[{Ahmed and Johnson(2021)}]{Ahmed_2021}
Ahmed, A.; and Johnson, F. 2021.
\newblock Gamification as a Way of Facilitating Emotions During
  Information-Seeking Behaviour: A Systematic Review of Previous Research.
\newblock In Toeppe, K.; Yan, H.; and Chu, S. K.~W., eds., \emph{Diversity,
  Divergence, Dialogue}, 85--98. Cham: Springer International Publishing.
\newblock ISBN 978-3-030-71305-8.

\bibitem[{Allam et~al.(2015)Allam, Kostova, Nakamoto, Schulz
  et~al.}]{Allam_2015}
Allam, A.; Kostova, Z.; Nakamoto, K.; Schulz, P.~J.; et~al. 2015.
\newblock The effect of social support features and gamification on a Web-based
  intervention for rheumatoid arthritis patients: randomized controlled trial.
\newblock \emph{Journal of medical Internet research}, 17(1): e3510.

\bibitem[{Armbruster and Klotzb\"{u}cher(2020)}]{Armbruster_2020}
Armbruster, S.; and Klotzb\"{u}cher, V. 2020.
\newblock Lost in lockdown? COVID-19, social distancing, and mental health in
  Germany.
\newblock Diskussionsbeitr\"{a}ge 2020-04, Freiburg i. Br.

\bibitem[{Bangor, Kortum, and Miller(2009)}]{Bangor_2009}
Bangor, A.; Kortum, P.; and Miller, J. 2009.
\newblock Determining What Individual SUS Scores Mean: Adding an Adjective
  Rating Scale.
\newblock \emph{J. Usability Stud.}, 4: 114--123.

\bibitem[{Brooke(1996)}]{Brooke_1996}
Brooke, J. 1996.
\newblock \emph{SUS -- a quick and dirty usability scale}, 189--194.

\bibitem[{Cellini et~al.(2020)Cellini, Canale, Mioni, and Costa}]{Cellini_2020}
Cellini, N.; Canale, N.; Mioni, G.; and Costa, S. 2020.
\newblock Changes in sleep pattern, sense of time and digital media use during
  COVID-19 lockdown in Italy.
\newblock \emph{Journal of Sleep Research}, 29(4): e13074.

\bibitem[{Ces\'{a}rio(2019)}]{Cesario_2019}
Ces\'{a}rio, V. 2019.
\newblock Guidelines for Combining Storytelling and Gamification: Which
  Features Would Teenagers Desire to Have a More Enjoyable Museum Experience?
\newblock In \emph{Extended Abstracts of the 2019 CHI Conference on Human
  Factors in Computing Systems}, CHI EA '19, 1–6. New York, NY, USA:
  Association for Computing Machinery.
\newblock ISBN 9781450359719.

\bibitem[{Clarke, DeNora, and Vuoskoski(2015)}]{Clarke_2015}
Clarke, E.; DeNora, T.; and Vuoskoski, J. 2015.
\newblock Music, empathy and cultural understanding.
\newblock \emph{Physics of Life Reviews}, 15: 61--88.

\bibitem[{Cuff et~al.(2016)Cuff, Brown, Taylor, and Howat}]{Cuff_2016}
Cuff, B.~M.; Brown, S.~J.; Taylor, L.; and Howat, D.~J. 2016.
\newblock Empathy: A Review of the Concept.
\newblock \emph{Emotion Review}, 8(2): 144--153.

\bibitem[{Elhadi et~al.(2021)Elhadi, Alsoufi, Msherghi, Alshareea, Ashini,
  Nagib, Abuzid, Abodabos, Alrifai, Gresea, Yahya, Ashour, Abomengal, Qarqab,
  Albibas, Anaiba, Idheiraj, Abraheem, Fayyad, Alkilani, Alsuwiyah, Elghezewi,
  and Zaid}]{Elhadi_2021}
Elhadi, M.; Alsoufi, A.; Msherghi, A.; Alshareea, E.; Ashini, A.; Nagib, T.;
  Abuzid, N.; Abodabos, S.; Alrifai, H.; Gresea, E.; Yahya, W.; Ashour, D.;
  Abomengal, S.; Qarqab, N.; Albibas, A.; Anaiba, M.; Idheiraj, H.; Abraheem,
  H.; Fayyad, M.; Alkilani, Y.; Alsuwiyah, S.; Elghezewi, A.; and Zaid, A.
  2021.
\newblock Psychological Health, Sleep Quality, Behavior, and Internet Use Among
  People During the COVID-19 Pandemic: A Cross-Sectional Study.
\newblock \emph{Frontiers in Psychiatry}, 12.

\bibitem[{Fiorillo et~al.(2020)Fiorillo, Sampogna, Giallonardo, Del~Vecchio,
  Luciano, Albert, Carmassi, Carrà, Cirulli, Dell’Osso, and
  et~al.}]{Fiorillo_2020}
Fiorillo, A.; Sampogna, G.; Giallonardo, V.; Del~Vecchio, V.; Luciano, M.;
  Albert, U.; Carmassi, C.; Carrà, G.; Cirulli, F.; Dell’Osso, B.; and
  et~al. 2020.
\newblock Effects of the lockdown on the mental health of the general
  population during the COVID-19 pandemic in Italy: Results from the COMET
  collaborative network.
\newblock \emph{European Psychiatry}, 63(1): e87.

\bibitem[{Friesem(2016)}]{Tettegah_2016}
Friesem, Y. 2016.
\newblock Chapter 2 - Empathy for the Digital Age: Using Video Production to
  Enhance Social, Emotional, and Cognitive Skills.
\newblock In Tettegah, S.~Y.; and Espelage, D.~L., eds., \emph{Emotions,
  Technology, and Behaviors}, Emotions and Technology, 21--45. San Diego:
  Academic Press.
\newblock ISBN 978-0-12-801873-6.

\bibitem[{Games(2022)}]{PeaceMaker}
Games, I. 2022.
\newblock PeaceMaker: Play the News, Solve the Puzzle.

\bibitem[{Gerber and Fischetti(2022)}]{Gerber_2022}
Gerber, A.; and Fischetti, B. 2022.
\newblock The Impact of Escape Room Gamification Using a Teleconferencing
  Platform on Pharmacy Student Learning.
\newblock \emph{Medical Science Educator}, 32.

\bibitem[{Giakalaras(2016)}]{Giakalaras_2016}
Giakalaras, M.~M. 2016.
\newblock Gamification and storytelling.
\newblock \emph{Univ. Aegean}, 8: 1--7.

\bibitem[{GOLDIE(2011)}]{Goldie_2011}
GOLDIE, P. 2011.
\newblock EMPATHY WITH ONE'S PAST.
\newblock \emph{The Southern Journal of Philosophy}, 49(s1): 193--207.

\bibitem[{Gómez-Urquiza et~al.(2019)Gómez-Urquiza, Gómez-Salgado,
  Albendín-García, Correa-Rodríguez, González-Jiménez, and {Cañadas-De la
  Fuente}}]{Gomez_2019}
Gómez-Urquiza, J.~L.; Gómez-Salgado, J.; Albendín-García, L.;
  Correa-Rodríguez, M.; González-Jiménez, E.; and {Cañadas-De la Fuente},
  G.~A. 2019.
\newblock The impact on nursing students' opinions and motivation of using a
  “Nursing Escape Room” as a teaching game: A descriptive study.
\newblock \emph{Nurse Education Today}, 72: 73--76.

\bibitem[{Hamari, Koivisto, and Sarsa(2014)}]{Hamari_2014}
Hamari, J.; Koivisto, J.; and Sarsa, H. 2014.
\newblock Does Gamification Work? -- A Literature Review of Empirical Studies
  on Gamification.
\newblock In \emph{2014 47th Hawaii International Conference on System
  Sciences}, 3025--3034.

\bibitem[{Hardee(2003)}]{Hardee_2003}
Hardee, J. 2003.
\newblock An Overview of Empathy.
\newblock \emph{The Permanente Journal}.

\bibitem[{Harteveld et~al.(2018)Harteveld, Snodgrass, Mohaddesi, Hart, Corwin,
  and Romera~Rodriguez}]{Harteveld_2018}
Harteveld, C.; Snodgrass, S.; Mohaddesi, O.; Hart, J.; Corwin, T.; and
  Romera~Rodriguez, G. 2018.
\newblock The Development of a Methodology for Gamifying Surveys.
\newblock In \emph{Proceedings of the 2018 Annual Symposium on Computer-Human
  Interaction in Play Companion Extended Abstracts}, CHI PLAY '18 Extended
  Abstracts, 461–467. New York, NY, USA: Association for Computing Machinery.
\newblock ISBN 9781450359689.

\bibitem[{Hung et~al.(2020)Hung, Lauren, Hon, Birmingham, Xu, Su, Hon, Park,
  Dang, and Lipsky}]{Hung_2020}
Hung, M.; Lauren, E.; Hon, E.~S.; Birmingham, W.~C.; Xu, J.; Su, S.; Hon,
  S.~D.; Park, J.; Dang, P.; and Lipsky, M.~S. 2020.
\newblock Social Network Analysis of COVID-19 Sentiments: Application of
  Artificial Intelligence.
\newblock \emph{J Med Internet Res}, 22(8): e22590.

\bibitem[{Huotari and Hamari(2017)}]{Huotari_2017}
Huotari, K.; and Hamari, J. 2017.
\newblock A definition for gamification: anchoring gamification in the service
  marketing literature.
\newblock \emph{Electronic Markets}, 27: 21--31.

\bibitem[{IJsselsteijn, {de Kort}, and Poels(2013)}]{IJsselsteijn_2013}
IJsselsteijn, W.; {de Kort}, Y.; and Poels, K. 2013.
\newblock \emph{The Game Experience Questionnaire}.
\newblock Technische Universiteit Eindhoven.

\bibitem[{Islam et~al.(2020)Islam, Bodrud-Doza, Khan, Haque, and
  Mamun}]{Islam_2020}
Islam, S. D.-U.; Bodrud-Doza, M.; Khan, R.~M.; Haque, M.~A.; and Mamun, M.~A.
  2020.
\newblock Exploring COVID-19 stress and its factors in Bangladesh: A
  perception-based study.
\newblock \emph{Heliyon}, 6(7): e04399.

\bibitem[{Junior et~al.(2021)Junior, Leite, Winum, Basso, Sousa, Nascimento,
  and Alves}]{Junior_2021}
Junior, J. N.~D.; Leite, A.; Winum, J.-Y.; Basso, A.; Sousa, U.; Nascimento,
  D.; and Alves, S. 2021.
\newblock HSG400 – Design, Implementation, and Evaluation of a Hybrid Board
  Game for Aiding Chemistry and Chemical Engineering Students in the Review of
  Stereochemistry During and After the COVID-19 Pandemic.
\newblock \emph{Education for Chemical Engineers}, 36.

\bibitem[{Kontoangelos, Economou, and Papageorgiou(2020)}]{Kontoangelos_2020}
Kontoangelos, K.; Economou, M.; and Papageorgiou, C. 2020.
\newblock Mental Health Effects of COVID-19 Pandemia: A Review of Clinical and
  Psychological Traits.
\newblock \emph{Psychiatry Investigation}, 17: 491--505.

\bibitem[{Kors et~al.(2016)Kors, Ferri, van~der Spek, Ketel, and
  Schouten}]{Kors_2016}
Kors, M.~J.; Ferri, G.; van~der Spek, E.~D.; Ketel, C.; and Schouten, B.~A.
  2016.
\newblock A Breathtaking Journey. On the Design of an Empathy-Arousing
  Mixed-Reality Game.
\newblock In \emph{Proceedings of the 2016 Annual Symposium on Computer-Human
  Interaction in Play}, CHI PLAY '16, 91–104. New York, NY, USA: Association
  for Computing Machinery.
\newblock ISBN 9781450344562.

\bibitem[{Lam and Tse(2022)}]{Lam_2022}
Lam, P.; and Tse, A. 2022.
\newblock Gamification in Everyday Classrooms: Observations From Schools in
  Hong Kong.
\newblock \emph{Frontiers in Education}, 6.

\bibitem[{Larcher et~al.(2021)Larcher, Pomatto, Battisti, Gullino, and
  Devecchi}]{Larcher_2021}
Larcher, F.; Pomatto, E.; Battisti, L.; Gullino, P.; and Devecchi, M. 2021.
\newblock Perceptions of Urban Green Areas during the Social Distancing Period
  for COVID-19 Containment in Italy.
\newblock \emph{Horticulturae}, 7(3).

\bibitem[{L\'{o}pez-Faican and Jaen(2021)}]{LopezFaican_2021}
L\'{o}pez-Faican, L.; and Jaen, J. 2021.
\newblock Designing Gamified Interactive Systems for Empathy Development.
\newblock In \emph{Companion Publication of the 2021 ACM Designing Interactive
  Systems Conference}, DIS '21 Companion, 27–29. New York, NY, USA:
  Association for Computing Machinery.
\newblock ISBN 9781450385596.

\bibitem[{making Lab(2022)}]{Go_Viral}
making Lab, S.~D. 2022.
\newblock Go Viral!

\bibitem[{Mauri et~al.(2022)Mauri, Tocchetti, Corti, Hsu, Verma, and
  Brambilla}]{Mauri_2022}
Mauri, A.; Tocchetti, A.; Corti, L.; Hsu, Y.-C.; Verma, H.; and Brambilla, M.
  2022.
\newblock COCTEAU: an Empathy-Based Tool for Decision-Making.

\bibitem[{Molnar(2018)}]{Molnar_2018}
Molnar, A. 2018.
\newblock The effect of interactive digital storytelling gamification on
  microbiology classroom interactions.
\newblock In \emph{2018 IEEE Integrated STEM Education Conference (ISEC)},
  243--246.

\bibitem[{Nieto-Escamez and Roldán-Tapia(2021)}]{Nieto_2021}
Nieto-Escamez, F.~A.; and Roldán-Tapia, M.~D. 2021.
\newblock Gamification as Online Teaching Strategy During COVID-19: A
  Mini-Review.
\newblock \emph{Frontiers in Psychology}, 12.

\bibitem[{Pantti and Tikka(2013)}]{Pantti_2013}
Pantti, M.; and Tikka, M. 2013.
\newblock \emph{Cosmopolitan empathy and user-generated disaster appeal videos
  on YouTube}.
\newblock International: Routledge.
\newblock ISBN 041581944X.

\bibitem[{Phan, Keebler, and Chaparro(2016)}]{Phan_2016}
Phan, M.; Keebler, J.; and Chaparro, B. 2016.
\newblock The Development and Validation of the Game User Experience
  Satisfaction Scale (GUESS).
\newblock \emph{Human Factors: The Journal of the Human Factors and Ergonomics
  Society}, 58.

\bibitem[{Philpot et~al.(2021)Philpot, Ramar, Roellinger, Barry, Sharma, and
  Ebbert}]{Philpot_2021}
Philpot, L.~M.; Ramar, P.; Roellinger, D.~L.; Barry, B.~A.; Sharma, P.; and
  Ebbert, J.~O. 2021.
\newblock Changes in social relationships during an initial "stay-at-home"
  phase of the COVID-19 pandemic: A longitudinal survey study in the U.S.
\newblock \emph{Soc Sci Med.}

\bibitem[{Rabbi et~al.(2017)Rabbi, Philyaw-Kotov, Lee, Mansour, Dent, Wang,
  Cunningham, Bonar, Nahum-Shani, Klasnja, Walton, and Murphy}]{Rabbi_2017}
Rabbi, M.; Philyaw-Kotov, M.; Lee, J.; Mansour, A.; Dent, L.; Wang, X.;
  Cunningham, R.; Bonar, E.; Nahum-Shani, I.; Klasnja, P.; Walton, M.; and
  Murphy, S. 2017.
\newblock SARA: A Mobile App to Engage Users in Health Data Collection.
\newblock In \emph{Proceedings of the 2017 ACM International Joint Conference
  on Pervasive and Ubiquitous Computing and Proceedings of the 2017 ACM
  International Symposium on Wearable Computers}, UbiComp '17, 781–789. New
  York, NY, USA: Association for Computing Machinery.
\newblock ISBN 9781450351904.

\bibitem[{Sauro(2011)}]{MeasuringUsability}
Sauro, J. 2011.
\newblock Measuring Usability with the System Usability Scale (SUS).

\bibitem[{Square(2022)}]{Coronavirus_Quiz}
Square, D.~P. 2022.
\newblock The Coronavirus Quiz.

\bibitem[{Steinmaurer, Sackl, and Gütl(2021)}]{Steinmaurer_2021}
Steinmaurer, A.; Sackl, M.; and Gütl, C. 2021.
\newblock Engagement in In-Game Questionnaires - Perspectives from Users and
  Experts.
\newblock In \emph{2021 7th International Conference of the Immersive Learning
  Research Network (iLRN)}, 1--7.

\bibitem[{Sánchez-Martín et~al.(2020)Sánchez-Martín, Corrales-Serrano,
  Luque-Sendra, and Zamora-Polo}]{Sanchez_2020}
Sánchez-Martín, J.; Corrales-Serrano, M.; Luque-Sendra, A.; and Zamora-Polo,
  F. 2020.
\newblock Exit for success. Gamifying science and technology for university
  students using escape-room. A preliminary approach.
\newblock \emph{Heliyon}, 6(7): e04340.

\bibitem[{Tan and Hsu(2023)}]{Tan_2023}
Tan, W.-K.; and Hsu, C.~Y. 2023.
\newblock The application of emotions, sharing motivations, and psychological
  distance in examining the intention to share COVID-19-related fake news.
\newblock \emph{Online Information Review}, 47(1): 59--80.

\bibitem[{Tocchetti and Brambilla(2020)}]{Tocchetti_2020}
Tocchetti, A.; and Brambilla, M. 2020.
\newblock A Gamified Crowdsourcing Framework for Data-Driven Co-Creation of
  Policy Making and Social Foresight.
\newblock In \emph{CSW@NeurIPS}.

\bibitem[{Tocchetti et~al.(2021)Tocchetti, Corti, Brambilla, and
  Marco}]{Tocchetti_2021}
Tocchetti, A.; Corti, L.; Brambilla, M.; and Marco, D.~D. 2021.
\newblock A Web-Based Co-Creation and User Engagement Method and Platform.
\newblock In \emph{International Conference on Web Engineering}.

\bibitem[{Triantoro et~al.(2020)Triantoro, Gopal, Benbunan-Fich, and
  Lang}]{Triantoro_2020}
Triantoro, T.; Gopal, R.; Benbunan-Fich, R.; and Lang, G. 2020.
\newblock Personality and games: enhancing online surveys through gamification.
\newblock \emph{Information Technology and Management}, 21.

\bibitem[{Videnovik et~al.(2022)Videnovik, Vold, Dimova, Ki{\o}nig, and
  Trajkovik}]{Videnovik_2022}
Videnovik, M.; Vold, T.; Dimova, G.; Ki{\o}nig, L.~V.; and Trajkovik, V. 2022.
\newblock Migration of an Escape Room--Style Educational Game to an Online
  Environment: Design Thinking Methodology.
\newblock \emph{JMIR Serious Games}, 10(3): e32095.

\bibitem[{White, Martin, and White(2022)}]{White_2022}
White, B.~K.; Martin, A.; and White, J. 2022.
\newblock Gamification and older adults: Opportunities for gamification to
  support health promotion initiatives for older adults in the context of
  COVID-19.
\newblock \emph{The Lancet Regional Health - Western Pacific}, 100528.

\bibitem[{Wright and McCarthy(2008)}]{Wright_2008}
Wright, P.; and McCarthy, J. 2008.
\newblock Empathy and Experience in HCI.
\newblock In \emph{Proceedings of the SIGCHI Conference on Human Factors in
  Computing Systems}, CHI '08, 637–646. New York, NY, USA: Association for
  Computing Machinery.
\newblock ISBN 9781605580111.

\end{thebibliography}

\end{document}